\documentclass{article}

    \PassOptionsToPackage{authoryear, compress, round}{natbib}
\usepackage[preprint]{neurips_2024}

\usepackage[utf8]{inputenc} % allow utf-8 input
\usepackage[T1]{fontenc}    % use 8-bit T1 fonts
\usepackage{hyperref}       % hyperlinks
\usepackage{url}            % simple URL typesetting
\usepackage{booktabs}       % professional-quality tables
\usepackage{amsfonts}       % blackboard math symbols
\usepackage{nicefrac}       % compact symbols for 1/2, etc.
\usepackage{microtype}      % microtypography
\usepackage{xcolor}         % colors
\usepackage{booktabs}
\usepackage{multirow}
\usepackage{graphicx}
\usepackage{colortbl}
\usepackage[shortlabels]{enumitem}
\usepackage{bbding}
\usepackage{caption}
\usepackage{subcaption}
\usepackage{wrapfig}
\usepackage{lipsum}
\usepackage{xspace}
\usepackage{multicol}
\usepackage{algorithm}
\usepackage{algpseudocode}
\usepackage{amsmath}
\usepackage{tcolorbox}
\usepackage{makecell}

\title{From MOOC to MAIC: Reshaping Online Teaching and Learning through LLM-driven Agents}

\author{%
  Jifan Yu$^1$, Zheyuan Zhang$^2$, Daniel Zhang-li$^2$, Shangqing Tu$^2$, Zhanxin Hao$^1$, \\ \textbf{Ruimiao Li$^1$, Haoxuan Li$^2$, Yuanchun Wang, Hanming Li$^2$, Linlu Gong$^2$, Jie Cao$^1$, } \\ \textbf{Jiayin Lin$^2$, Jinchang Zhou$^2$, Fei Qin$^1$, Haohua Wang$^2$, Jianxiao Jiang$^1$, Lijun Deng$^2$, } \\ \textbf{Yisi Zhan$^1$, Chaojun Xiao$^2$, Xusheng Dai$^2$, Xuan Yan$^2$,  Nianyi Lin$^1$, } \\ \textbf{Nan Zhang$^3$, Ruixin Ni$^3$, Yang Dang$^1$, Lei Hou$^2$, Yu Zhang$^1$, Xu Han$^2$,} \\ \textbf{ Manli Li$^1$, Juanzi Li$^2$, Zhiyuan Liu$^2$\thanks{Corresponding Authors.}, Huiqin Liu$^1$\footnotemark[1], Maosong Sun$^2$\footnotemark[1]}\\
  $^1$Institute of Education, Tsinghua University.\\ 
  $^2$Department of Computer Science and Technology, Tsinghua University.\\ 
  $^3$ModelBest Inc. \\
  \texttt{thu\_maic@tsinghua.edu.cn} \\
}

\begin{document}

\maketitle

\begin{abstract}
Since the first instances of online education, where courses were uploaded to accessible and shared online platforms, this form of scaling the dissemination of human knowledge to reach a broader audience has sparked extensive discussion and widespread adoption. Recognizing that personalized learning still holds significant potential for improvement, new AI technologies have been continuously integrated into this learning format, resulting in a variety of educational AI applications such as educational recommendation and intelligent tutoring. The emergence of intelligence in large language models (LLMs) has allowed for these educational enhancements to be built upon a unified foundational model, enabling deeper integration. In this context, we propose MAIC (Massive AI-empowered Course), a new form of online education that leverages LLM-driven multi-agent systems to construct an AI-augmented classroom, balancing scalability with adaptivity. Beyond exploring the conceptual framework and technical innovations, we conduct preliminary experiments at Tsinghua University, one of China’s leading universities. Drawing from over $100,000$ learning records of more than $500$ students\footnote{We follow the approval from Tsinghua University Sci.\&Tech. Ethics Committee (NO.THU-04-2024-56).}, we obtain a series of valuable observations and initial analyses. This project will continue to evolve, ultimately aiming to establish a comprehensive open platform that supports and unifies research, technology, and applications in exploring the possibilities of online education in the era of large model AI. We envision this platform as a collaborative hub, bringing together educators, researchers, and innovators to collectively explore the future of AI-driven online education.
\end{abstract}

\section{Introduction}

\begin{figure}[h]
    \centering
    \includegraphics[width=\linewidth]{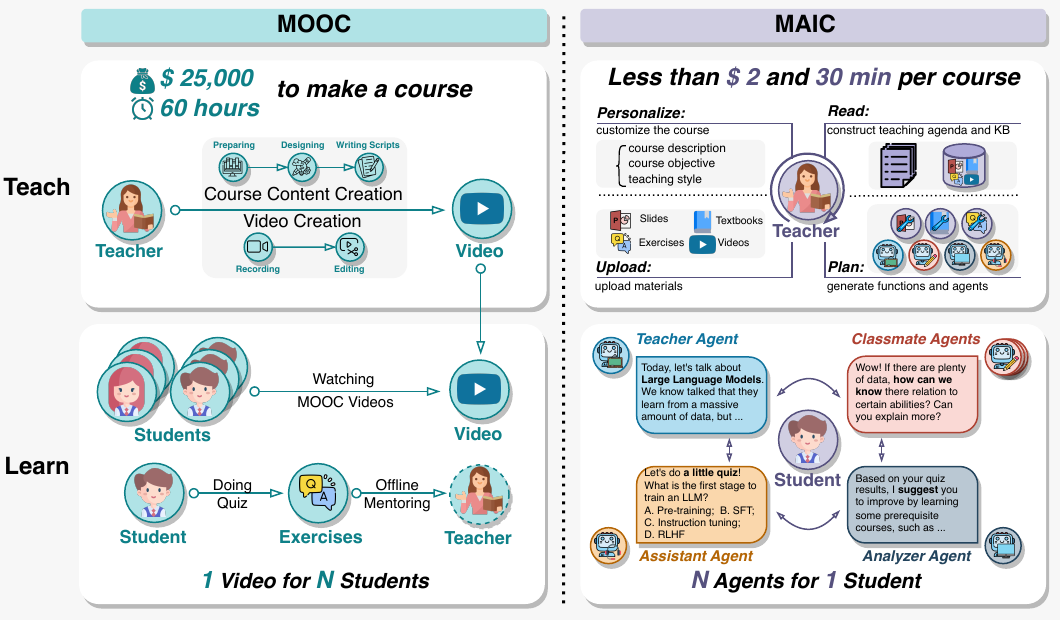}
    \caption{MOOC V.S. MAIC from the aspects of teaching and learning.}
    \label{fig:MOOC2MAIC}
\end{figure}

\textbf{Explicit Background: Evolution for \textit{Scalability}}. The evolution of online education stands as a testament to humanity's relentless pursuit of knowledge, transcending the limitations of time and space~\citep{koller2013online}. From the humble beginnings of oral tradition to the advent of the printed book~\citep{froebel1886education,halstead2005values}, education has continually sought ways to expand its reach. Yet, for centuries, the traditional model of education was bound by the constraints of physical classrooms, limited resources, and localized instruction. The dawn of the internet marked a revolutionary shift, heralding the age of online education, where the dream of universal access to knowledge began to take tangible form. Specifically, the Massive Open Online Course (MOOC) phenomenon marks a significant milestone in the evolution of online education, reflecting both technological advancement and educational innovation~\citep{MOOC, daniel2012making}. 
Since then, platforms like edX\footnote{\url{https://www.edx.org/}}, involving institutions such as MIT and Harvard, and Coursera\footnote{\url{https://www.coursera.org/}}, originating from Stanford, have integrated learning resources from over $270$ renowned universities~\citep{pappano2012year,papadakis2023moocs}. These platforms have attracted more than $100$ million learners globally, progressively realizing the \textit{\textbf{Scalability}} of online education.

\textbf{Implicit Motivation: Determination of \textit{Adaptivity}}. However, this paradigm of serving thousands of learners from diverse backgrounds through one pre-recorded video~\citep{reich2019mooc} (as shown in Figure \ref{fig:MOOC2MAIC}) struggles to align with the educational philosophy of ``teaching in accordance with individual aptitudes''~\citep{reich2015rebooting,zhu2020comprehensive}. This challenge has become a significant reason for the subsequent efforts that introducing of AI techniques into online learning. To achieve the \textbf{\textit{Adaptivity}} of learning, a series of tasks such as learning path planning~\citep{nabizadeh2020learning,zhong2022towards}, course recommendation~\citep{jiang2019goal,zhang2018mcrs,jing2017guess}, and intelligent tutoring~\citep{yilmaz2022smart,tu2023littlemu}—driven by technologies like recommendation systems and dialogue generation—have been employed to enhance the student learning experience. 

% \begin{figure}
%     \centering
%     \includegraphics[width=0.5\linewidth]{figures/MOOC_and_MAIC.pdf}
%     \caption{Caption}
%     \label{fig:mooc-vs-maic}
% \end{figure}

Although these technologies have been applied across various aspects of teaching and learning, the significant differences among the supporting tasks before, during, and after instruction have posed challenges for unifying them under a single deep learning framework~\citep{kabudi2021ai}. Such fragmentation has, in part, delayed the emergence of a new platform where AI and online learning are fully integrated. However, with the rapid advancement of generative AI~\citep{epstein2023art}, large language models (LLMs) have created fresh opportunities for AI-powered learning paradigms. Models such as GPT-4~\citep{achiam2023gpt} and LLaMA~\citep{touvron2023llama} possess strong generalization capabilities and encapsulate vast parametric knowledge, allowing for the flexible configuration of intelligent agents~\citep{chen2023agentverse} built upon them. Currently, LLM-driven multi-agent systems~\citep{wu2023autogen} have already been explored for applications such as social simulation~\citep{park2023generative} and the execution of complex tasks like software development~\citep{qian2023communicative}. This progress opens up a potential pathway for introducing large language model multi-agent systems to create entirely new online teaching and learning experiences.

\textbf{Proposal of MAIC}. At the critical juncture of a new era defined by large language models and multi-agent systems in online education, we introduce MAIC (Massive AI-empowered Course). MAIC is dedicated to exploring the integration of multi-agent systems across various stages of online learning, including course preparation, instruction, and analysis, with the goal of balancing \textbf{\textit{Scalability}} and \textbf{\textit{Adaptivity}} of online education. The core concept of MAIC is to construct a series of LLM-driven agents to support both \textbf{Teaching} and \textbf{Learning} in the online educational environment. 
As illustrated in the Figure~\ref{fig:MOOC2MAIC}, the paradigm of MOOC and MAIC can be featured with two primary aspects:

\textbf{Teaching}: This action is primarily performed by the instructor. In previous MOOC, the instructor is responsible for thoroughly preparing course materials, drafting lecture notes, and spending considerable time meticulously recording the course. The final output typically consists of a series of pre-recorded instructional videos. For the proposed MAIC, however, the instructor only needs to upload the teaching slides. With intelligent assistance, the instructor can further complete the PPT creation, after which the agents, utilizing a range of models such as multimodal understanding and knowledge structure extraction, will generate structured lecture notes and learning resources optimized for use by the AI system.
\textbf{Learning}: In MOOCs~\citep{reich2019mooc}, a single set of course materials is designed to serve thousands of students with diverse backgrounds, and the pace of instruction is predetermined by the instructor, offering very limited room for personalized adaptation based on individual student needs~\citep{yu2021mooccubex}. While in MAIC, course delivery is autonomously managed by AI teacher agents, which dynamically adjust the teaching process based on student interactions and inquiries. Additionally, MAIC offers AI teaching assistants and customizable AI classmates. Students can select the AI agents they wish to study with, thereby creating varied classroom scenarios that provide personalized learning companions, emotional support, and opportunities for knowledge discussion.

% \textbf{Primary }

In this report, we introduce the MAIC platform, offering an intuitive and user-friendly solution that accommodates the needs of various users, including students and educators. This platform comes pre-equipped with a suite of intelligent agents and tools that support course analysis and the construction of new MAIC course examples. Additionally, MAIC integrates several learning analytics tools powered by large models, enabling quick access to learning data, forecasting academic outcomes, and automating tasks such as interviews and assessments.

With the support of Tsinghua University, one of China's top universities, we conducted an exploration of this new learning model over a period of more than three months. Assisted by over $500$ student volunteers, we implemented the study using two courses: the AI course ``Towards Artificial General Intelligence'' (TAGI) and the learning science course ``How to Study in the University'' (HSU). During this pilot, we collected over $100,000$ behavioral records. Based on the data from these courses, along with student survey measurements and qualitative interview results, we conducted an initial analysis of the features and performance of the MAIC system. In subsequent sections, we will briefly introduce the technical implementation, algorithm improvement and primary results of MAIC~\footnote{Our open-source Demo and detailed analysis will be released soon. More technical details are introduced in the companion papers Slide2Lecture and SimClass~\citep{zhang2024simulating}.}. 

\section{Related Work}

\paragraph{\textbf{AI-assisted Online Learning}} 
Online learning~\citep{pappano2012year} refers to the process of acquiring knowledge within an electronic environment composed of communication technologies, network infrastructure, artificial intelligence, and multimedia tools. While online education significantly enhances learners’ access to knowledge, supports personalized learning, and facilitates learning tailored to individual needs, persistent issues such as low course completion rates and suboptimal learning outcomes remain formidable challenges~\citep{zhong2022towards}. A retrospective study published by MIT~\citep{reich2019mooc} highlights that the lack of continuous guidance and personalized support for online learners is a critical factor affecting the quality of learning and student development in online settings. The inherent nature of online and remote learning often results in physical separation between students and instructors, making it prohibitively expensive to maintain real-time interactions through manual means~\citep{aleven2015beginning}. Therefore, AI researchers have begun to introduce auxiliary learning applications such as resource recommendation systems~\citep{jing2017guess} and intelligent teaching assistants~\citep{jiang2019massistant,tu2023littlemu} into online education environments. Leveraging technologies like educational knowledge graphs~\citep{dang2019mooc,yu2020mooccube}, they are increasingly integrating diverse technologies to construct personalized online learning systems that enhance the learning experience through tailored support and adaptive learning pathways. In the era of large language models, platforms like Khan Academy have pioneered the deployment of AI-driven tools, such as the Khanmigo virtual tutor~\footnote{\url{https://blog.khanacademy.org/teacher-khanmigo/}}. This development has sparked discussions among researchers about the potential to move beyond basic question-answering and recommendation functionalities, exploring the design of more deeply integrated models that fuse AI with online education in innovative ways~\citep{golchin2024large}. Some of the research efforts gradually aim to create new paradigms that go beyond traditional approaches to AI-enhanced learning.

\paragraph{\textbf{LLM-driven AI Tutoring System}} The evolution of intelligent tutoring systems (ITS) from early expert systems to agent-based models in the era of large language models has undergone several transformations in interaction paradigms~\citep{chowdhury_autotutor_2024}. During the 1980s and 1990s, ITS began leveraging expert systems and related technologies to deliver instruction aligned with learners' cognitive processes across diverse educational settings. Representative systems from this period include AutoTutor, developed by the University of Memphis~\citep{nye2014autotutor}, and SCOT from Stanford University[51]. These systems offered greater flexibility and were among the first to support natural language-based question-and-answer interactions. However, the content of these interactions remained pre-designed, limiting their ability to provide learning support beyond the scope of the system's initial design. The advent of Large Language Models (LLMs) has significantly broadened the scope of intelligent tutoring systems (ITS), offering unprecedented interactivity and adaptability across educational platforms~\citep{bubeck2023sparks}. Recent strides in the study of multi-agent systems and the integration of tools have catalyzed new approaches to planning and student interaction within ITS frameworks~\citep{Park2023GenerativeAgents, qian2023communicative, schick2024toolformer}. Emerging research~\citep{chen_empowering_2023} illustrates the capacity of LLMs to autonomously curate and deliver educational content by leveraging an array of tools. Additionally, innovations like MWPTutor~\citep{chowdhury_autotutor_2024} explore how LLMs can effectively manage the teaching of complex subjects, such as mathematical word problems, demonstrating their versatility in specialized learning environments. Building on these developments, our research focuses on an LLM-based ITS model designed to establish a robust framework for comprehensive, lecture-level tutoring that aligns with evolving educational standards, thereby reinforcing the transformative role of LLMs in the future of education.

\section{MAIC}

% \subsection{Overview}
In this section, we introduce the key techniques for implementing a MAIC platform. Specifically, we present the main workflows designed for both the teaching and learning sides, highlighting the key challenges associated with each and the corresponding solutions implemented to address these issues.

\subsection{Teaching: Course Preparation Workflow}
\label{sec:teach}

\begin{figure}
    \centering
    \includegraphics[width=\linewidth]{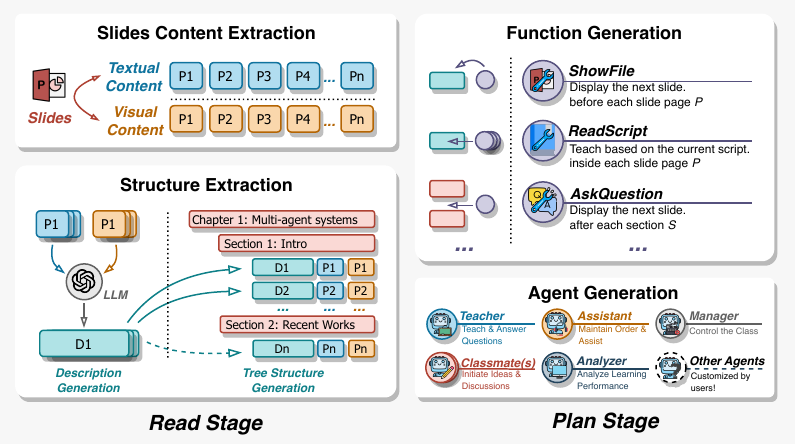}
    \caption{An illustration of the course preparation workflow of MAIC.}
    \label{fig:MAIC-pre}
\end{figure}

To transform vast amounts of weakly structured and static learning resources~\citep{dang2019mooc} into highly structured and adaptive learning materials~\citep{yu2021mooccubex}, MAIC develop a standardized course preparation workflow. This workflow is designed to streamline the workload of experts, facilitating the scalability of this online learning model and preparing it for broader implementation. Such course preparation workflow of MAIC consists of two major stage: \textbf{Read} and \textbf{Plan}. 

\subsubsection{Read Stage}

At this stage, instructors (and authorized teaching assistants) are involved by providing material. With the assistance of multi-agent systems empowered by large language models, they upload a set of course slides $\mathcal{P}=\left\{P_i \right\}^{1\leq i\leq |\mathcal{P}|}$, which are then transformed into highly structured intelligent learning resources $ \widehat{\mathcal{P}}=\left\{\left<P_i,D_i,K_j \right> \right\}^{1\leq i\leq |\mathcal{P}|}_{1\leq j\leq |\mathcal{P}|}$ along with multiple AI agents designed for classroom construction. The $P_i$ and $D_i$ here correspond to a single slide page and its textual description, while $K_j$ denotes to the knowledge-aware section of each page.

\textbf{1. Slides Content Extraction}. First, MAIC employs a multi-modal LLM (mLLM)~\citep{huang2023chatgpt} to capture the textual content $P_{i}^{t}$ and the visual content $P_{i}^{v}$ of each page of the given slide $\mathcal{P}$, i.e., $f_{T}^{1}: P_{i} \rightarrow <P_{i}^{t} , P_{i}^{v}>$. Such functional model $f_{T}^{1}$~\footnote{The same applies to similar symbols in the following text.} can be adjusted and further improved by the arising LLM techniques, and the current implementation is based on the GPT-4V~\footnote{\url{https://openai.com/index/gpt-4v-system-card/}} model with certain prompting contexts.

\textbf{2. Structure Extraction}. After the pre-processing of the uploaded slides, MAIC employs two functions to complete the read stage. The produced $<P_{i}^{t} , P_{i}^{v}>$ are described by an mLLM-based method with straightforward and comprehensive texts, i.e., $f_{T}^{2}: <P_{i}^{t} , P_{i}^{v}> \rightarrow D_{i}$. Meanwhile, MAIC takes an knowledge extraction method to organize the core knowledge of each page and build a tree-style taxonomy for the slide, i.e., $f_{T}^{3}: <P_{i}^{t} , P_{i}^{v}, D_{i}> \rightarrow {K_j}$, which makes up the final $\widehat{\mathcal{P}}$. 

% In current implementation, 

\subsubsection{Plan Stage}
\label{sec:plan}

At this stage, instructors (and authorized teaching assistants) are involved by proofreading and refining the results. Based on highly structured slides, MAIC constructs a novel instructional action representation language, allowing the incorporation of flexible teaching functions such as lecturing and questioning into preset classrooms, which naturally connects with related educational technologies like lecture script generation~\citep{hong2023visual} and question generation~\citep{kurdi2020systematic}. Meanwhile, leveraging intelligent agent construction techniques~\citep{chen2023agentverse}, the platform utilizes course content to provide teachers and teaching assistants with AI-driven agents, facilitating the early planning of the foundational structure for subsequent courses. 

\textbf{3. Function Generation.} To make the heterogeneous teaching actions be generated within the classroom context, teaching activities such as lecturing and giving quizzes are conceptualized as \textit{teaching actions} within MAIC. Each teaching action $\mathcal{T}$ is defined as $\mathcal{T} = (type, value)$, where $type$ indicates the category of the action (e.g., $ShowFile$, $ReadScript$, $AskQuestion$), and $value$ details the content of the action, such as the script to be read aloud. This approach reflects our principles of flexibility and adaptability, allowing classroom actions to be easily configurable. It empowers developers and educators to create custom teaching actions tailored to specific needs, facilitating their smooth integration into the overall teaching process.

Each Function is associated with the generation of certain content, denoted as $<\mathcal{T}_{n}, \widehat{P}_{\mathcal{T}}>$. For instance, the function $AskQuestion$ necessitates linkage with one or a set of specific questions. Among these, the most crucial action is $ReadScript$, as it constitutes the core of the instructional process. Based on this function, these Teaching Actions are embedded within the course script using special marker symbols, thereby enabling the intelligent agent to read and invoke them in a personalized manner. Specifically, MAIC has trained a high-quality lecture script generation model based on long-context encoding methods and multi-modal model foundations, thereby supporting the fundamental class procedures as well as the integration of other teaching actions, i.e., $f_{T}^{4}: \widehat{P} \rightarrow \widehat{P}_{script}$. Then, MAIC provide a series of optional functions such as $f_{T}^{5}: \widehat{P} \rightarrow \widehat{P}_{question}$ for serving proactive questioning during the class. Note that all the generated results are required to be checked and adjusted by instructors, which guarantees the quality and correctness of the produced course.

\textbf{4. Agent Generation.} Meanwhile, instructors can provide personalized information (such as the voice, teaching styles, and extended course material) for building customized teaching agents, such as Teacher Agent $a_{T}$ and Teaching Assistant Agent $a_{TA}$. MAIC provides several agentization toolkits~\citep{chen2023agentverse} that implemented via LLMs, supporting the high-level DIY of these agents. The extended course materials uploaded are also segmented in this section and integrated into different intelligent agents using the RAG (Retrieval-Augmented Generation) technology. The associated series of technological innovations are also thoroughly introduced and evaluated in the concurrent academic papers.

\subsection{Learning: Multi-agent Classroom Environment}

As described in the Introduction, student learning in MAIC follows a \textit{"1 Student user + N AI Agents"} model. In such an environment, the AI teacher controls the learning progress based on the highly structured instructional action representation language mentioned earlier, explains course content, poses questions, and navigates PowerPoint slides, while the AI teaching assistant maintains classroom order and prevents content deviation. Students can interrupt the teacher at any time, ask questions, and engage in discussions, and the intelligent agents continuously adjust the teaching process and some content based on the students' performance. As introduce in~\citet{zhang2024simulating}, the design principles for constructing this immersive adaptive classroom originate from the following two concerns: (1) How to ensure that the classroom covers the core classroom behaviors? (2) How to maintain the entirety of the interaction within the natural flow of the classroom process? 

\begin{figure}
    \centering
    \includegraphics[width=\linewidth]{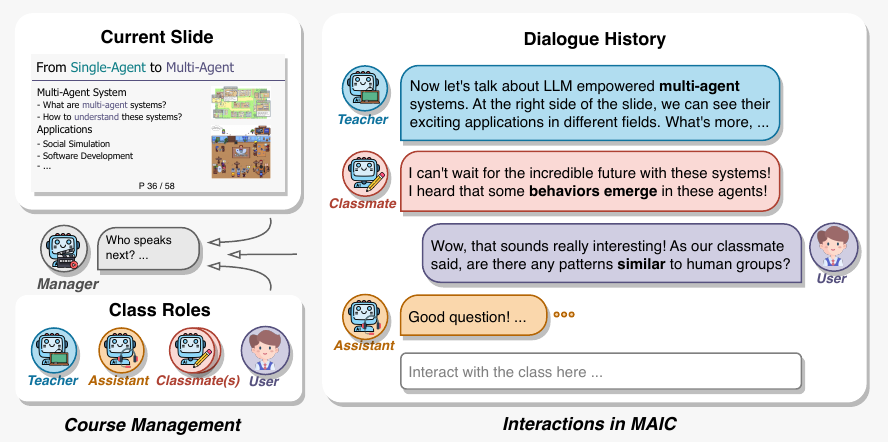}
    \caption{An illustration of the classroom learning environment of MAIC.}
    \label{fig:MAIC-class}
\end{figure}

In addressing the initial concern, we systematically classify classroom interaction behaviors in accordance with established educational principles, as delineated in Schwanke's seminal work~\citep{schwanke1981classroom}: \textit{Teaching and Initiation (TI)} encompasses the instructive actions of the teacher and the responsive feedback or insights provided by students; \textit{In-depth Discussion (ID)} involves the alignment, deliberation, and iterative question-and-answer exchanges between the teacher and students, which are instrumental in facilitating students' conceptual comprehension; \textit{Emotional Companionship (EC)} pertains to the encouragement of student learning, the cultivation of a conducive learning environment, and the provision of emotional sustenance; and \textit{Classroom Management (CM)} refers to the maintenance of order, the organization of disruptive elements, and the steering of classroom discourse. Recognizing that these pedagogical behaviors manifest through diverse \textbf{Class Roles} (represented as $ \widehat{ \mathcal{R}}= \left\{ r_{i} \right\}^{\left| \widehat{ \mathcal{R}} \right|}$, with each $r{i}$ signifying a distinct role), it is imperative to ensure the \textit{variety} and \textit{breadth} of the agents' roles within the educational setting.

Addressing the subsequent concern, we emphasize the necessity of meticulously and rhythmically orchestrating the interactions among the various agents within the system, in harmony with the course content. With the Learning Materials (denoted as $C = \left [ c_1,...,c_t \right ]$, where each instructional script $c_t$ is sequenced), we introduce an innovative \textbf{Session Controller} designed to regulate the flow of classroom interactions, contingent upon the class's dynamic state and under the aegis of a central managerial agent~\citep{wu2023autogen}.

Based on these principles, we construct multiple classmate agents for diverse roles, implement class control, and ultimately derive the multi-agent classroom process.

\textbf{Classmate Agents.} To enhance the educational experience and emulate the dynamics of traditional classroom settings, we currently preset a variety of student-like agents, each imbued with unique personality traits, to complement the teaching agents. These agents are designed to perform roles akin to peer students, enriching the interactive landscape of the learning environment. In this scholarly work, we have introduced an initial set of four archetypal student agents, while also providing users with the flexibility to customize and introduce additional engaging student agents onto the educational platform. Each agent $\mathbf{a}_{i} \in \mathcal{A}$ is facilitated through prompting LLMs and associated with one or more class roles, denoted as:
     $\mathcal{A} = \rho \left ( LLM, \mathsf{P}_{A} \right ), \mathcal{A} \Leftrightarrow \widehat{\mathcal{R}}$, 
where $\rho$ is the role customization operation, $\mathsf{P}_{A}$ is the system prompt with agent description~\citep{zhang2024simulating}.

$\bullet$ \textit{Class Clown (TI, EC, CM)}: Crafted to spark creativity, engender a lively classroom ambiance, and act as a supportive peer, this agent also assists the teacher in steering the class's focus when the learner's attention wanders.

$\bullet$ \textit{Deep Thinker (TI, ID)}: This agent is dedicated to profound contemplation and to posing thought-provoking questions that challenge and extend the intellectual boundaries of the classroom.

$\bullet$ \textit{Note Taker (TI, CM)}: With a penchant for summarizing and disseminating key points from the class discussions, this agent aids in the cognitive organization and retention of information for all participants.

$\bullet$ \textit{Inquisitive Mind (TI, EC)}: Characterized by a propensity for inquiring about lecture content, this agent fosters a culture of inquiry and dialogue, prompting others to engage in critical thinking and collaborative discourse.

Unlike Standardized Operating Procedures (SOPs) commonly used in multi-agent systems~\citep{qian2023communicative, hong2023metagpt}, classroom scenarios function as dynamic, interactive environments without rigid workflows, resembling an evolving group discussion. In these settings, agents must determine appropriate timing for their interactions, adapting to the fluid nature of classroom discourse. To address this need, we designed a controller that observes classroom dynamics, makes informed decisions, and manages agents’ behaviors based on the current state of the class. The Session Controller is composed of three core modules: the Class State Receptor and Manager Agent.

\textbf{Class State Receptor.} The Class State Receptor captures the ongoing classroom dialogue, with the history up to time $t$ represented as $H_t = \bigcup (u_i^{\mathbf{a}_j})^{t}$, where $u_i$ is the utterance made by agent $\mathbf{a}j$ or a user (denoted as $\mathbf{a}{u}$). The class state $S_t$ integrates this interaction data, structured as $\mathcal{S}_t = \left\{P_t, H_t | \widehat{\mathcal{R}}   \right\}$. Here, $P_t \subseteq P$ represents the learning materials covered up to time $t$. This design prioritizes adaptability and real-time decision-making, aligning with pedagogical principles that emphasize responsiveness to the evolving needs of learners within an educational setting.

\textbf{Manager Agent.} Drawing inspiration from AutoGen~\citep{wu2023autogen} and MathVC~\citep{yue2024mathvc}, we designed a hidden meta-agent responsible for regulating the dynamics of classroom interactions. This agent receives the current class state $\mathcal{S}_t$, monitors the flow of the class, interprets ongoing activities, and determines the subsequent action to be executed, ensuring that the learning environment remains adaptive and responsive. The task $f_\mathcal{L}$ of the Manager Agent can be formally defined as $f_\mathcal{L}: \mathcal{S}_t \rightarrow \left ( \mathbf{a}_t, \mathcal{T} \right ) | \mathbf{a}_t \in \mathcal{A}, \mathcal{T}_{n} \Leftarrow \mathcal{T} $.

where $\mathcal{T}_n$ denotes a specific function, and the selected action will be carried out, transitioning the class to the next state. After executing an action, the system enters a waiting phase for a time window $\tau$. During this period, if a user responds or the waiting time elapses, the Manager Agent is triggered to make a new decision. This design reflects key educational principles by prioritizing a learner-centered approach, maintaining fluid class engagement, and promoting timely and contextually relevant instructional adjustments, thereby enhancing the overall educational experience.

This classroom management method is the core of MAIC learning stage. Currently, we collect plenty interaction data and employ several foundation models~\citep{hu2024minicpm, glm2024chatglm} via fine-tuning or prompt tuning as our baseline model.

\section{Key Technique Evaluation}

MAIC is a complex large language model-based intelligent agent system that encompasses various specific technologies. On the teaching side, it involves multiple processes for content generation and knowledge understanding, while on the learning side, it requires evaluating the effectiveness of agent construction and classroom management capabilities. Currently, we focus on presenting two core functions: lecture script generation and course management, which are fundamental to the teaching and learning aspects of MAIC. The evaluation of other technologies will be continuously updated. It is important to note that these assessments provide only an initial view of specific aspects of MAIC rather than its overall effectiveness, which will be further explored through real-world practice and results analysis in subsequent sections.

\subsection{Teaching Side Evaluation}

\textbf{Function Generation}. As introduced in Section~\ref{sec:teach}, generating vivid slide scripts is the core function of MAIC teaching workflow. \textbf{Baselines}: To assess the effectiveness of our implementation, we established two baseline configurations for comparison: (1) We replicated \textit{\textbf{S}cript\textbf{2}\textbf{T}ranscript}~\citep{nguyen_automatic_nodate}, referred to as \textit{S2T}, which uses slide titles to offer overarching context and guidance for generating transcripts. (2) We also reproduced \textit{\textbf{S}elf-\textbf{C}ritique \textbf{P}rompting}~\citep{olney_automatic_2024}, denoted as \textit{SCP}, which incorporates a self-critique and refinement process to enhance script quality.

\textbf{Evaluation Metrics}: Our evaluation employs four distinct metrics to rate the generated scripts on a 5-point Likert scale, where 1 represents unacceptable quality and 5 denotes optimal performance:
\begin{enumerate}
    \item \textit{Tone} evaluates whether the script appropriately reflects the instructional tone of a teacher.
    \item \textit{Clarity} measures how clear and comprehensible the script is for learners.
    \item \textit{Supportiveness} assesses the extent to which the script provides emotional and motivational support to students.
    \item \textit{Alignment} evaluates the degree to which the script content aligns with the slide material.
\end{enumerate}

Overall performance is determined by averaging the scores across all metrics.

\textbf{Procedure}: We executed the course preparation pipeline for each baseline and collected script evaluations from annotators. To minimize potential bias, each slide was assessed by three independent annotators, who were required to provide ratings for all configurations of the same slide. This approach ensures a balanced and comprehensive evaluation of the teaching material, aligning with educational principles that emphasize clarity, support, and contextual relevance in instructional content.

\begin{table}[th]
    \centering
    \small
    \caption{Evaluation result of our implementation of generation.}
    \begin{tabular}{l|cccc|c}
        \toprule
        \textbf{Setting} & \textbf{Tone} & \textbf{Clarity} & \textbf{Supportive} & \textbf{Matching} & \textbf{Overall}\\
        \midrule
        S2T~\citep{nguyen_automatic_nodate} & 3.88 & 3.93 & 3.23 & 3.63 & 3.67\\
        SCP~\citep{olney_automatic_2024} & \textbf{4.03} & \underline{4.24} & 3.38 & 3.93 & \underline{3.90}\\
        \midrule
         MAIC-FuncGen & 4.00 & \textbf{4.25} & \textbf{3.57} & \textbf{4.18} & \textbf{4.00}\\
         w/o visual & 3.78 & 3.73 & \underline{3.44} & 3.51 & 3.61\\
         w/o context & 3.97 & 4.00 & 3.38 & \underline{4.03} & 3.84\\
        \midrule
        Human & \underline{4.02} & 4.07 & 3.38 & 3.98 & 3.86\\
        \bottomrule
    \end{tabular}
    \label{tab:teacher-readscript}
\end{table}

\textbf{Results}: As shown in Table~\ref{tab:teacher-readscript}, our approach (MAIC-FuncGen) achieves the highest overall score of $4.00$, outperforming all baseline methods. Our analysis reveals several key insights:

\begin{itemize}
    \item \textbf{Importance of Visual Input}: The inclusion of visual elements significantly enhances script generation quality. Both \textit{S2T} and MAIC-FuncGen without visual inputs received lower matching scores, highlighting the need for contextual visual cues that align the script content closely with the presented materials.
    
    \item \textbf{Role of Contextual Information}: The presence of coherent contextual information, including content from previous pages, is critical for script quality. It not only improves the clarity of the current content but also enhances the supportive and matching aspects by providing a continuous and interconnected learning narrative. This aligns with pedagogical principles that emphasize coherence and context in learning materials to support deeper understanding.
    
    \item \textbf{Comparative Performance with Human Instructors}: Interestingly, our approach slightly outperformed the human baseline across three key dimensions. This can be attributed to the inherent ability of large language models (LLMs) to strictly adhere to instructions, maintaining alignment with slide content and employing an encouraging and supportive tone. In contrast, human instructors often expand on topics freely, introducing their own style and diverging from the core content, which reflects a more flexible, albeit less structured, instructional approach.
\end{itemize}

These findings underscore the value of integrating structured visual and contextual information into script generation, while also highlighting the potential of LLMs to complement traditional instructional strategies by providing consistency and structured support.

\subsection{Learning Side Evaluation}

\begin{wrapfigure}{r}{0.5\linewidth}        
    \includegraphics[width=\linewidth]{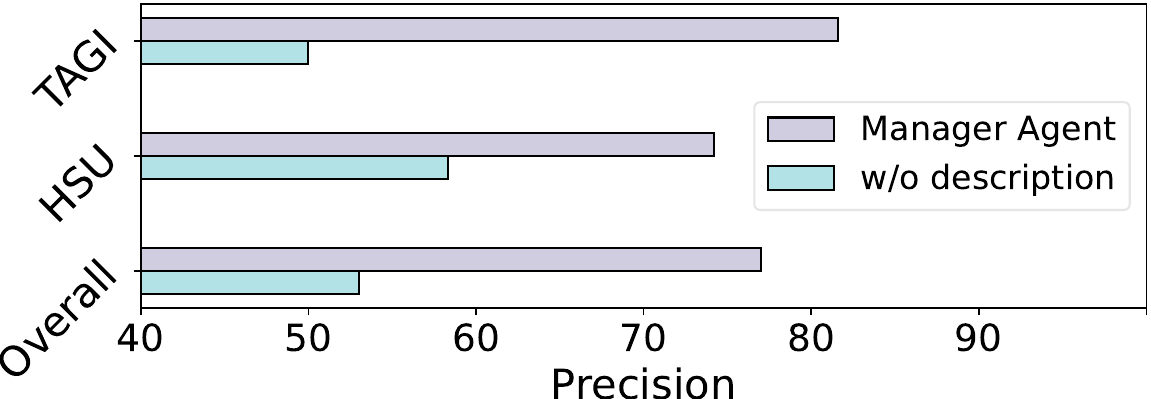} 
    \caption{Manager Agent Precision.}
    \label{tec:manager}
\end{wrapfigure} 
\textbf{Classroom Manager Agent}. Section~\ref{sec:plan} also mention several relevant techniques of MAIC learning. The classroom manager agent is the keypoint of the classroom controlling. \textbf{Setup:} However, the evaluation of this process is highly subjective, making it challenging to establish an objective scoring system for assessment. Therefore, in the practical implementation of the TAGI and HSU courses, we select $500$ actual system decisions and extracted their corresponding classroom scenarios. We recruit expert teachers and teaching assistants to manually annotate these scenarios. Based on this annotated data, we derive the results shown in Figure~\ref{tec:manager}. These results illustrate the alignment between the actions chosen by the manager agent and those selected by human instructors in determining the next course action. Specifically, we evaluate the implementation with and without role description, detecting the effects of these contextual information.

\textbf{Result:} Statistical analysis reveals that omitting role descriptions for each agent reduces the classifier's performance. Although the LLM can sometimes identify the correct agent by referencing partial behaviors from the chat history, the inclusion of comprehensive role descriptions markedly enhances performance. This suggests that while leveraging chat history as input for the scene controller can provide some benefits, it is insufficient for consistently generating accurate outputs.

The current results, however, remain below optimal levels, indicating further opportunities to refine and enhance the user experience. Despite the suboptimal performance, interacting agents demonstrate the capacity to partially offset these shortcomings. This compensatory effect is due to the LLM's ability to manage user queries beyond the predefined functions of each agent, as evidenced by our subsequent behavioral study, where user ratings did not significantly decline in the ablation setting.

Nonetheless, enhancing the accuracy of the controller agent remains advantageous, as agents can more effectively manage tasks they are specifically designed for. For example, the teacher agent is tailored to adopt a softer, more instructive tone, but it may be less effective in handling safety-related cases compared to the teaching assistant agent. Improved accuracy ensures that each agent operates within its designed scope, contributing to a more seamless and effective instructional process.

\section{Behavioral Experiment}

Following approval from Tsinghua University Science and Technology Ethics Committee (Certificate No: THU-04-2024-56) and the recruitment of student and teacher volunteers, we conduct over three months of teaching practice and behavioral analysis in the courses "Towards General Artificial Intelligence" and "How to Study in the University." This large-scale study involves more than 500 students and aims to address three core questions: \textbf{Q1:} What is the quality of MAIC Courses? \textbf{Q2:} What are the learning outcomes within MAIC? \textbf{Q3:} How do students perform in the MAIC environment? In the following sections, we present some preliminary observations from this study.

\subsection{Q1: The Quality of MAIC Course}

We evaluated the quality of the MAIC course using the results from two questionnaires completed by course takers. The first questionnaire focused on the quality of AI teacher's teaching, adapted from the Community of Inquiry Framework~\citep{garrison2007researching}. Items in the original COI questionnaire were revised to make them suitable for the AI-engaged learning environment. This questionnaire was administered when students completed the whole course. 

\begin{wrapfigure}{r}{0.5\linewidth}        
    \includegraphics[width=\linewidth]{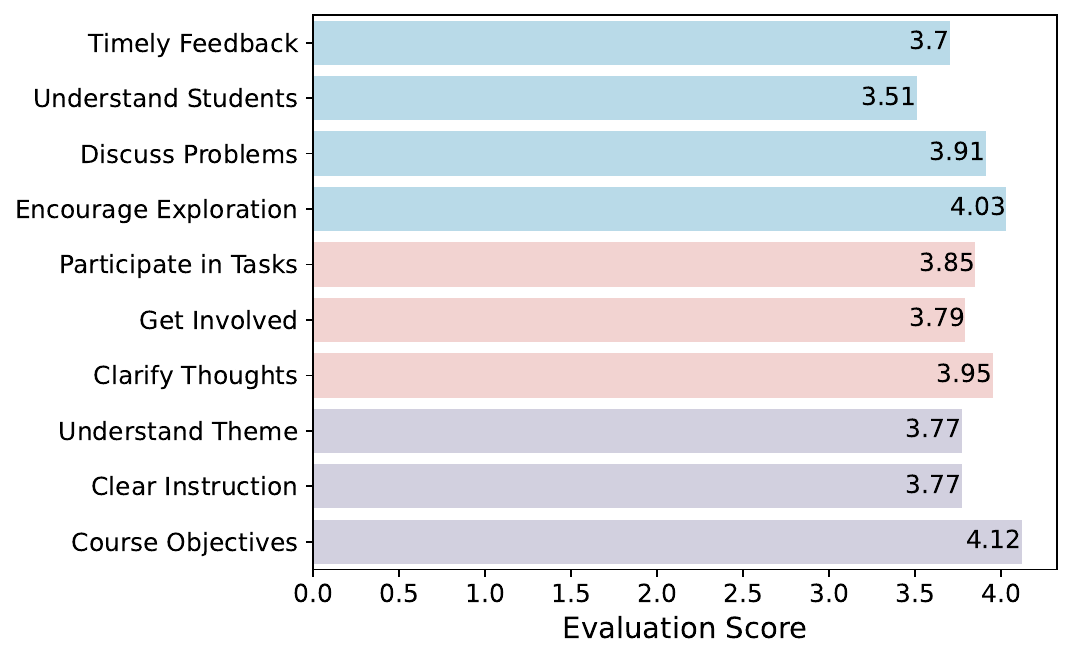} 
    \caption{Results from after-course survey.}
    \label{beh:feedback}
\end{wrapfigure} 

As presented in Figure~\ref{beh:feedback}, the results showed that overall students had positive beliefs of the teaching quality on MAIC. For instance, the mean score of students' rating on the question "The AI instructor clearly communicated important course goals" (Course Objective) was 4.12 (SD = $0.66$) out of $5$, and the average rating on "The AI instructor encouraged course participants to explore new concepts" was $4.03$ (SD = $0.73$). These findings suggest that students felt the AI instructors effectively helped them understand course objectives, clarify their thinking, explore new ideas, and engage in meaningful dialogue.

However, relatively lower ratings were observed on questions like ``The AI instructor provided feedback that helped me understand my strengths and weaknesses'' (Understand Student), which had a mean score of $3.51$ (SD = $0.94$). This indicates that AI instructors may lack personalization and adaptability during the teaching process, possibly because the same scripts were used for all students.

\subsection{Q2: The Behavior of MAIC Student Engagement}

Firstly, when it comes to choosing the class mode, students tend to prefer the "continuous mode," believing that this mode allows them to maintain their train of thought without interruption, thereby ensuring learning efficiency. The "continuous mode" refers to a setting where, after selecting the teacher and other intelligent agent roles (such as teaching assistant, thinker, note-taker), the chosen roles conduct the class from start to finish without any interactive input from the students, making it a relatively passive learning approach. For example, during interviews, one student mentioned: 

\begin{center}
\fcolorbox{black}{gray!10}{\parbox{.9\linewidth}{"I mostly used the continuous mode because, in the interactive mode, after the AI teacher finishes each sentence, you have to respond before they can continue. I don't always feel like interacting after every sentence, so most of the time, I use continuous mode. Of course, there were one or two times when I used interactive mode because the AI prompted me to speak, but I remember that once, after I spoke in interactive mode, the AI didn't respond or react to what I said, so I felt like it wasn't very useful. After that, I just stuck with the continuous mode." In this mode, although interaction is not possible, some students adopt a pause strategy if they don't understand something. For instance, one student mentioned, "If I didn't understand something, I would pause and look at the PPT and the text in the text box. I don't think I ever stopped to ask the AI to explain something again, unless it was some unfamiliar term or a more exploratory topic."}}
\end{center}

Secondly, regarding specific behaviors during the class, some students proactively ask questions to the intelligent agent around certain topics. As shown in Figure 3, $61$\% of the students' behavior in class involved actively seeking knowledge, information, or asking questions. For example, asking "Can you explain the transformer structure in simple terms?" The interview results of this study also support this view. Many students indicated that they would actively ask questions. Some examples from the interviews include:
\begin{center}
\fcolorbox{black}{gray!10}{\parbox{.9\linewidth}{$\bullet$  "When I asked a question, the AI would ask a follow-up question related to mine, and I felt it was an amazing experience, like it was really guiding me to think more deeply. I think the questions it asked made a lot of sense. In that interactive mode, it could extend into many other discussions beyond the course content." \\
$\bullet$  "I think this might be an advantage of AI teaching. Because, in a traditional classroom, whether it's a large class or even a small one, students nowadays are generally reluctant to ask questions. There are various reasons for this. Also, I think immediate Q\&A helps me learn the material better. If I have a question, I can get an answer right away, and I think that immediate feedback is really valuable."}}
\end{center}
\begin{wrapfigure}{r}{0.5\linewidth}        
    \includegraphics[width=\linewidth]{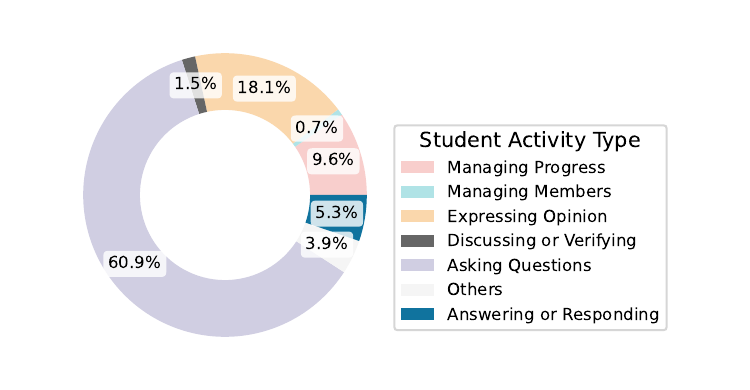} 
    \caption{Ratio of student activities.}
    \label{beh:stu_act}
\end{wrapfigure} 
Additionally, some students manage and control different intelligent agent roles and the class progress. These management-related behaviors account for $11$\% (Figure~\ref{beh:stu_act}). For example, "Please go back to the previous slide," or "Please explain that in simpler terms," demonstrating strong autonomy and self-regulated learning abilities. In the interviews, some students also mentioned, \textit{I don't just ask for knowledge; I might ask, 'I want to learn more,' or 'I hope to explore something new in a certain field,' to manage and regulate the AI's responses.}

Overall, the findings reveal that while students prefer the "continuous mode" for its uninterrupted flow of information, this passive approach may limit opportunities for active engagement and critical thinking. In the future, AI agent-driven classes should actively encourage student interaction rather than simply delivering continuous knowledge. In addition, the high level of proactive questioning indicates that students are eager to engage when given the opportunity, underscoring the importance of designing AI tools that foster inquiry-based learning~\citep{xie2023influence}. Questioning is an important behavior that reflects active learning in students, which ultimately leads to better academic performance. Finally, the occurrence of management-related behaviors suggests that students already realize their active role in their learning process, which align with \citet{cain2024prompting}' study. It is said that the capabilities of LLM AI tools can help students and educators transition from passive recipients to active co-creators of their learning experiences. Further research can give more support and encouragement for them to take an active role in directing their learning.

\subsection{Q3: The Outcome of MAIC Learning}

We assessed the effectiveness of the MAIC course from three perspectives: performance in module tests and the final exam, technology acceptance through questionnaires, and self-reported higher-order thinking scores.

\textbf{Test Results.} Module tests were conducted at the end of each module, focusing primarily on the content covered in the most recent module. The final exam was administered one week after the course concluded, synthesizing questions from the earlier module tests. Average attendance of the module tests is $76.3\%$ (SD=$6\%$), while that of the final exam is $73.3\%$. Test scores (stantardized to percentage) ranged from $53.3\%$ (Module 2, SD=$18.9\%$) to $82.4\%$ (Module 4, SD=$16.3\%$), reflecting students' learning outcomes. These outcomes were corroborated by student interviews. One participant expressed, ``My most impressive gains, from the perspective of knowledge, all came from the post-class test. If there were no post-class test, I might not remember the knowledge at all, and it might just pass by like a passing cloud of smoke... When I took the test and then looked back at the courseware, it was the peak period for me to absorb knowledge, honestly speaking.''
\begin{wraptable}{r}{0.6\textwidth}
    \centering
    \caption{Correlation of students' message-aware behavior and test results. Values shown in the table are normalized.}
    \label{tab:message_beh}
    \begin{tabular}{l|rr}
\toprule
w/o control & $\mu(log(\text{MsgNum}))$    & $\mu(log(\text{MsgLen}))$  \\ \midrule
AvgQuiz & 0.341***    & 0.202* \\
FinalExam & 0.346***   & 0.333** \\ \midrule
w/ control & $\mu(log(\text{MsgNum}))$    & $\mu(log(\text{MsgLen}))$  \\ \midrule
AvgQuiz & 0.206**    & 0.177** \\
FinalExam & 0.174   & 0.235* \\ \bottomrule
\end{tabular}
\end{wraptable}

Additionally, test scores are strongly associated with class engagement. Specifically, the frequency (measured by the logorithm of the number of messages per module) and length of in-class chat messages (measured by the logorithm of the number of characters per message and module) —a prominent feature of the MAIC system—were positively correlated with standardized test scores and final exam performance, as presented in Table~\ref{tab:message_beh}. A regression analysis of standardized test scores on in-class chat message metrics was performed, controlling for normalized scores from Module 1. As Table~\ref{tab:message_beh} shows, in-class chat engagement is found to significantly predict higher test scores.

\textbf{Technology Acceptance.} We further evaluate the acceptance of generative AI tools, such as ChatGPT, before and after the course. The results demonstrated a significant overall increase in technology acceptance ($N=111$, $t=3.05$, $p=0.002$). Further analysis of specific dimensions of acceptance revealed significant improvements in Habit ($t=2.81$, $p=0.005$), Effort Expectancy ($t=3.98$, $p<0.001$), and Facilitating Conditions ($t=3.22$, $p=0.002$). These findings indicate that students grew increasingly accustomed to and supportive of the MAIC course format. 
Interview responses also reflected an enhanced understanding and acceptance of AI technologies. One student remarked:

\begin{center}
\fcolorbox{black}{gray!10}{\parbox{.9\linewidth}{"I used to be quite resistant to AI, mainly because it was too complicated, but after this course, I found that it was not that complicated, and I became more accepting of it. I also learned some of its principles."}}
\end{center}

\textbf{High-order Thinking}. We explore the perceived impact of using LLMs on students' higher-order thinking skills through pre- and post-course questionnaires. Comparative analyses and t-test results showed significant increases in students' perceptions of the positive effects of the course on their abstract thinking ($t=2.32$, $p=0.02$) and critical thinking ($t=2.37$, $p=0.02$). These findings suggest that students believe using large language models throughout the course enhanced their cognitive abilities in these areas.Interviews further illuminated these perceptions. One student stated:  

\begin{center}
\fcolorbox{black}{gray!10}{\parbox{.9\linewidth}{"I think it might make me more confident in asking these questions and think more," while another mentioned, "Unlike before, (I was always) hiding and waiting for the opportunity to ask again."}}
\end{center}

However, the course's impact on other higher-order thinking skills remained ambiguous. Several students noted a lack of deep thinking and discussion opportunities in the course. For instance, one student commented:

\begin{center}
\fcolorbox{black}{gray!10}{\parbox{.9\linewidth}{"(In a real classroom) After class, I can ask the teacher to explain it to me again... At this time, the teacher will definitely give you different ideas or explanations, but in this class, you will have no more after class." Another student added, "this course is purely about content because the teacher's teaching is quite mechanical. The course is rich in theoretical knowledge, but there is almost no life thinking, life enlightenment, or some enlightenment and thinking outside of artificial intelligence."}}
\end{center}

Notably, there is a limitation came from the questionnaire itself. As we evaluated students' perceptions on the impacts on high-order thinking abilities, it should be noted that the abilities are not estimated. Future study should include designed scales or tasks to estimate the effects of the course on the students' high-order thinking abilities. 

To address this limitation, future research should incorporate well-designed scales or specific tasks that can objectively measure students' higher-order thinking abilities. These might include assessments that evaluate critical thinking, problem-solving, and analytical reasoning skills directly. By doing so, researchers can gain a more comprehensive understanding of the course's effectiveness in enhancing these abilities. Moreover, the inclusion of such measures would allow for a more rigorous evaluation of the course's impact, providing stronger evidence to support or challenge the findings based on student perceptions alone.

\section{Predicted Impact and Ethical Consideration}
\subsection{Predicted Impact}
The implementation of MAIC in online education is expected to revolutionize the learning experience by enhancing both scalability and adaptability. By leveraging multi-agent systems, MAIC can dynamically adjust to the needs of individual learners, providing personalized learning paths that were previously unattainable in traditional MOOCs. This personalized approach not only improves learning outcomes but also provides access to high-quality education across diverse socio-economic backgrounds, as it reduces the dependency on human instructors for content delivery.

Moreover, MAIC is anticipated to address some of the inherent challenges of traditional online education, such as the one-size-fits-all model and the lack of real-time adaptability~\citep{rizvi2022beyond}. The integration of AI-driven agents as teachers, teaching assistants, and classmates creates a more interactive and responsive learning environment. This shift promises to increase engagement and motivation among students, potentially leading to higher completion rates and deeper understanding of the material.

However, it is crucial to acknowledge that the introduction of such a transformative system could also have unintended consequences. There may be a widening gap between students who adapt well to AI-powered learning environments and those who struggle with this new mode of education. Additionally, the reliance on AI systems could lead to reduced opportunities for human instructors, potentially diminishing the role of educators in the learning process. These impacts need to be carefully monitored and addressed through ongoing evaluation and refinement of the MAIC system.
\subsection{Ethical Considerations}
The deployment of MAIC in online education brings forth significant ethical considerations that must be carefully evaluated~\citep{almarzouqi2024ethical}. One of the foremost concerns is learner privacy and data security. The system's reliance on large-scale data collection and analysis to personalize learning experiences raises questions about how student data is stored, accessed, and used. To mitigate these concerns, stringent data protection measures have been implemented, including encryption and anonymization of student records. However, given the sensitivity of educational data, continuous efforts to enhance security protocols are essential.

Issues of discrimination and bias also present ethical challenges. While MAIC aims to provide a personalized learning experience for all students, there is a risk that the algorithms driving these personalized experiences may inadvertently reinforce existing biases, particularly if the training data is not sufficiently diverse. To address this, the development team has incorporated fairness-focused auditing procedures into the algorithmic design process. Although these measures are designed to minimize bias, it is recognized that no system is entirely immune to these challenges, and ongoing monitoring is required to ensure equitable outcomes.

The accuracy of information and content regulation within MAIC is another critical area of ethical concern. As the system automates the creation and dissemination of educational content, there is a risk that inaccuracies could be propagated at scale. To mitigate this, content generated by the AI is regularly reviewed by subject matter experts and teaching assistants. Nonetheless, given the vast scale of content production, it is acknowledged that some errors may still occur. Thus, the system includes mechanisms for students and educators to flag and correct inaccuracies, ensuring continuous improvement of the educational material.

In terms of the ethics of education, MAIC raises questions about the role of teachers in a system that increasingly relies on AI-driven instruction. While the system enhances scalability and provides personalized learning experiences, it also diminishes the direct involvement of human educators. This shift may impact the development of teacher-student relationships, which are critical for fostering emotional and social growth in learners~\citep{thornberg2022teacher}. To address this, MAIC includes human-in-the-loop design principles, ensuring that educators can intervene and guide the AI’s decision-making processes when necessary. However, the balance between AI automation and human oversight remains a complex issue that requires ongoing consideration.

Furthermore, the lack of peer interaction in a predominantly AI-driven educational environment could hinder the development of important social skills among students. To counter this, MAIC incorporates AI classmates designed to simulate peer interactions. While these agents provide a form of interaction, they cannot fully replicate the nuances of human-to-human communication. The system, therefore, encourages mixed-mode learning environments where students can engage with both AI and human peers, preserving the benefits of social learning.

Finally, the issue of student personalization must be approached with caution. While MAIC’s ability to adapt to individual learning needs is a significant advantage, it also poses risks to fairness and equality. There is a potential for certain students to receive more tailored and effective instruction based on their data profiles, potentially exacerbating educational inequalities. To mitigate this, MAIC includes mechanisms to ensure that all students, regardless of their data profiles, have access to high-quality learning experiences. The system’s fairness algorithms are continuously refined to promote equitable educational outcomes for all learners.

\section{Conclusion}

In this paper, we provide a concise overview of the development trajectory of online education and the technological opportunities arising in the era of large language models. Considering the principles of adaptivity and scalability, along with the sophisticated design of LLM-driven multi-agent systems, we explore how existing MOOC can be transformed into MAIC (Massive AI-empowered Courses) and discuss new paradigms of teaching and learning. We propose a comprehensive solution, analyze the key technical components, and implement each step of the process. Our approach was practically deployed in two courses at Tsinghua University, leading to a series of preliminary observations of student behavior. These initial findings suggest that highly personalized classrooms built with new AI-assisted learning technologies can achieve high quality, and student behavior demonstrates the effectiveness of the teaching process. Moving forward, this work will be continuously maintained and expanded, aiming to develop an open and shared platform for educational exploration, academic research, and technological innovation. We hope our work will call upon and serve educational theorists, technology developers, and innovators to engage in discussions about the new online environment in the era of large language models.

\section*{Acknowledgement}

\subsection{Author Contribution}

\textbf{System Implementation}. Jifan Yu, Zheyuan Zhang, and Daniel Zhang-li designed the overall framework of the system. Zheyuan Zhang refined the workflow representation method used for controlling agents, while Daniel Zhang-li was responsible for the development and engineering deployment of several algorithms, including resource processing and function generation. Shangqing Tu oversaw security reviews and the integration of RAG methods. Linlu Gong collected the evaluation data from students. Nan Zhang and Ruixin Ni were responsible for project management and coordination throughout the development phase of MAIC, playing a crucial role in ensuring the quality of the final system. 

\textbf{Theoretical Investigation}. Zhanxin Hao and Ruimiao Li were responsible for the theoretical investigation and analysis of the MAIC concept, while Yang Dang contributed to the early development of this idea. All supervising professors provided valuable insights and guidance in the conceptualization of MAIC.

\textbf{Toolkit Completion}. Haoxuan Li, Yuanchun Wang, Hanming Li, Jiayin Lin, Jinchang Zhou, and Nianyi Lin contributed to the development of various MAIC tools, including the cognitive diagnosis module, automated interview module, and automatic analysis module. Haohua Wang and Lijun Deng played significant roles in data collection and processing. 

\textbf{Course Practice}. Yisi Zhan and Chaojun Xiao were instrumental in the first round of MAIC pilot courses, handling a wide range of practical tasks, including the recruitment of student volunteers, provision of course materials, content review, and post-class management. Xusheng Dai refined the course practice process, while Xuan Yan was deeply involved in course support activities. Their efforts were critical to the successful execution of the two courses.

\textbf{Pedagogical analysis}. Under the guidance of Yu Zhang, Zhanxin Hao was responsible for the preliminary pedagogical evaluation of the MAIC system. Ruimiao Li, under the supervision of Manli Li, conducted key field research. Jie Cao, Fei Qin, and Jianxiao Jiang played crucial roles in the analysis process, overseeing tasks such as automated coding, qualitative interviews, and data analysis, respectively.

\textbf{Paper Writing}. Jifan Yu was responsible for the primary writing of the manuscript, while Zhanxin Hao and Ruimiao Li contributed significantly to the design and execution of the behavioral experiments and the ethical consideration.

\textbf{Advising}. Manli Li, Juanzi Li, Zhiyuan Liu, Huiqin Liu, and Maosong Sun took advisor roles in this project. Xu Han brought technical insights into the system design. This project also received guidance and support from various relevant departments at Tsinghua University.

\subsection{Acknowledgement}

This research project is supported by a grant from the Institute for Guo Qiang, Tsinghua University (20192920479). 

This project would like to express its appreciation for the contributions and assistance of many other participants. The artistic design was skillfully provided by Shanshan Wang. The platform development and feature implementation were carefully carried out by Peng Zhou, Yuting Liu, Yuanwei Xu and Chengqiang Xu.  

\bibliographystyle{plainnat}
\bibliography{neurips_2024}

\begin{thebibliography}{54}
\providecommand{\natexlab}[1]{#1}
\providecommand{\url}[1]{\texttt{#1}}
\expandafter\ifx\csname urlstyle\endcsname\relax
  \providecommand{\doi}[1]{doi: #1}\else
  \providecommand{\doi}{doi: \begingroup \urlstyle{rm}\Url}\fi

\bibitem[Achiam et~al.(2023)Achiam, Adler, Agarwal, Ahmad, Akkaya, Aleman, Almeida, Altenschmidt, Altman, Anadkat, et~al.]{achiam2023gpt}
Josh Achiam, Steven Adler, Sandhini Agarwal, Lama Ahmad, Ilge Akkaya, Florencia~Leoni Aleman, Diogo Almeida, Janko Altenschmidt, Sam Altman, Shyamal Anadkat, et~al.
\newblock Gpt-4 technical report.
\newblock \emph{arXiv preprint arXiv:2303.08774}, 2023.

\bibitem[Aleven et~al.(2015)Aleven, Sewall, Popescu, Xhakaj, Chand, Baker, Wang, Siemens, Ros{\'e}, and Gasevic]{aleven2015beginning}
Vincent Aleven, Jonathan Sewall, Octav Popescu, Franceska Xhakaj, Dhruv Chand, Ryan Baker, Yuan Wang, George Siemens, Carolyn Ros{\'e}, and Dragan Gasevic.
\newblock The beginning of a beautiful friendship? intelligent tutoring systems and moocs.
\newblock In \emph{Artificial Intelligence in Education: 17th International Conference, AIED 2015, Madrid, Spain, June 22-26, 2015. Proceedings 17}, pages 525--528. Springer, 2015.

\bibitem[Almarzouqi et~al.(2024)Almarzouqi, Aburayya, Alfaisal, Elbadawi, and Salloum]{almarzouqi2024ethical}
Amina Almarzouqi, Ahmad Aburayya, Raghad Alfaisal, Mohamed~Ahmad Elbadawi, and Said~A Salloum.
\newblock Ethical implications of using chatgpt in educational environments: A comprehensive review.
\newblock \emph{Artificial Intelligence in Education: The Power and Dangers of ChatGPT in the Classroom}, pages 185--199, 2024.

\bibitem[Bubeck et~al.(2023)Bubeck, Chandrasekaran, Eldan, Gehrke, Horvitz, Kamar, Lee, Lee, Li, Lundberg, Nori, Palangi, Ribeiro, and Zhang]{bubeck2023sparks}
Sébastien Bubeck, Varun Chandrasekaran, Ronen Eldan, Johannes Gehrke, Eric Horvitz, Ece Kamar, Peter Lee, Yin~Tat Lee, Yuanzhi Li, Scott Lundberg, Harsha Nori, Hamid Palangi, Marco~Tulio Ribeiro, and Yi~Zhang.
\newblock Sparks of artificial general intelligence: Early experiments with gpt-4, 2023.
\newblock URL \url{https://arxiv.org/abs/2303.12712}.

\bibitem[Cain(2024)]{cain2024prompting}
William Cain.
\newblock Prompting change: exploring prompt engineering in large language model ai and its potential to transform education.
\newblock \emph{TechTrends}, 68\penalty0 (1):\penalty0 47--57, 2024.

\bibitem[Chen et~al.(2023)Chen, Su, Zuo, Yang, Yuan, Qian, Chan, Qin, Lu, Xie, et~al.]{chen2023agentverse}
Weize Chen, Yusheng Su, Jingwei Zuo, Cheng Yang, Chenfei Yuan, Chen Qian, Chi-Min Chan, Yujia Qin, Yaxi Lu, Ruobing Xie, et~al.
\newblock Agentverse: Facilitating multi-agent collaboration and exploring emergent behaviors in agents.
\newblock \emph{arXiv preprint arXiv:2308.10848}, 2\penalty0 (4):\penalty0 6, 2023.

\bibitem[Chen et~al.(2024)Chen, Ding, Zheng, Liu, Sun, and Zhou]{chen_empowering_2023}
Yulin Chen, Ning Ding, Hai-Tao Zheng, Zhiyuan Liu, Maosong Sun, and Bowen Zhou.
\newblock Empowering private tutoring by chaining large language models, 2024.
\newblock URL \url{https://arxiv.org/abs/2309.08112}.

\bibitem[Cormier(2008)]{MOOC}
D~Cormier.
\newblock The cck08 mooc – connectivism course, 1/4 way, 2008.
\newblock URL \url{https://davecormier.com/edblog/2008/10/02/the-cck08-mooc-connectivism-course-14-way/}.

\bibitem[Dang et~al.(2019)Dang, Tang, and Li]{dang2019mooc}
Furong Dang, Jintao Tang, and Shasha Li.
\newblock Mooc-kg: A mooc knowledge graph for cross-platform online learning resources.
\newblock In \emph{2019 IEEE 9th International Conference on Electronics Information and Emergency Communication (ICEIEC)}, pages 1--8. IEEE, 2019.

\bibitem[Daniel(2012)]{daniel2012making}
John Daniel.
\newblock Making sense of moocs: Musings in a maze of myth, paradox and possibility.
\newblock \emph{Journal of interactive Media in education}, 2012\penalty0 (3):\penalty0 18--18, 2012.

\bibitem[Epstein et~al.(2023)Epstein, Hertzmann, of~Human~Creativity, Akten, Farid, Fjeld, Frank, Groh, Herman, Leach, et~al.]{epstein2023art}
Ziv Epstein, Aaron Hertzmann, Investigators of~Human~Creativity, Memo Akten, Hany Farid, Jessica Fjeld, Morgan~R Frank, Matthew Groh, Laura Herman, Neil Leach, et~al.
\newblock Art and the science of generative ai.
\newblock \emph{Science}, 380\penalty0 (6650):\penalty0 1110--1111, 2023.

\bibitem[Froebel(1886)]{froebel1886education}
Friedrich Froebel.
\newblock \emph{The education of man}, volume~5.
\newblock A. Lovell \& Company, 1886.

\bibitem[Garrison and Arbaugh(2007)]{garrison2007researching}
D~Randy Garrison and J~Ben Arbaugh.
\newblock Researching the community of inquiry framework: Review, issues, and future directions.
\newblock \emph{The Internet and higher education}, 10\penalty0 (3):\penalty0 157--172, 2007.

\bibitem[GLM et~al.(2024)GLM, Zeng, Xu, Wang, Zhang, Yin, Rojas, Feng, Zhao, Lai, et~al.]{glm2024chatglm}
Team GLM, Aohan Zeng, Bin Xu, Bowen Wang, Chenhui Zhang, Da~Yin, Diego Rojas, Guanyu Feng, Hanlin Zhao, Hanyu Lai, et~al.
\newblock Chatglm: A family of large language models from glm-130b to glm-4 all tools.
\newblock \emph{arXiv preprint arXiv:2406.12793}, 2024.

\bibitem[Golchin et~al.(2024)Golchin, Garuda, Impey, and Wenger]{golchin2024large}
Shahriar Golchin, Nikhil Garuda, Christopher Impey, and Matthew Wenger.
\newblock Large language models as moocs graders.
\newblock \emph{arXiv preprint arXiv:2402.03776}, 2024.

\bibitem[Halstead and Taylor(2005)]{halstead2005values}
Mark Halstead and Monica~J Taylor.
\newblock \emph{Values in education and education in values}.
\newblock Routledge, 2005.

\bibitem[Hong et~al.(2023{\natexlab{a}})Hong, Zheng, Chen, Cheng, Wang, Zhang, Wang, Yau, Lin, Zhou, et~al.]{hong2023metagpt}
Sirui Hong, Xiawu Zheng, Jonathan Chen, Yuheng Cheng, Jinlin Wang, Ceyao Zhang, Zili Wang, Steven Ka~Shing Yau, Zijuan Lin, Liyang Zhou, et~al.
\newblock Metagpt: Meta programming for multi-agent collaborative framework.
\newblock \emph{arXiv preprint arXiv:2308.00352}, 2023{\natexlab{a}}.

\bibitem[Hong et~al.(2023{\natexlab{b}})Hong, Sayeed, Mehra, Demberg, and Schiele]{hong2023visual}
Xudong Hong, Asad Sayeed, Khushboo Mehra, Vera Demberg, and Bernt Schiele.
\newblock Visual writing prompts: Character-grounded story generation with curated image sequences.
\newblock \emph{Transactions of the Association for Computational Linguistics}, 11:\penalty0 565--581, 2023{\natexlab{b}}.

\bibitem[Hu et~al.(2024)Hu, Tu, Han, He, Cui, Long, Zheng, Fang, Huang, Zhao, et~al.]{hu2024minicpm}
Shengding Hu, Yuge Tu, Xu~Han, Chaoqun He, Ganqu Cui, Xiang Long, Zhi Zheng, Yewei Fang, Yuxiang Huang, Weilin Zhao, et~al.
\newblock Minicpm: Unveiling the potential of small language models with scalable training strategies.
\newblock \emph{arXiv preprint arXiv:2404.06395}, 2024.

\bibitem[Huang et~al.(2023)Huang, Zheng, Wang, Yin, Wang, Ding, Yin, Xu, Yang, Zheng, et~al.]{huang2023chatgpt}
Hanyao Huang, Ou~Zheng, Dongdong Wang, Jiayi Yin, Zijin Wang, Shengxuan Ding, Heng Yin, Chuan Xu, Renjie Yang, Qian Zheng, et~al.
\newblock Chatgpt for shaping the future of dentistry: the potential of multi-modal large language model.
\newblock \emph{International Journal of Oral Science}, 15\penalty0 (1):\penalty0 29, 2023.

\bibitem[Jiang et~al.(2019{\natexlab{a}})Jiang, Hu, Huang, Wang, Yang, Ye, and Zheng]{jiang2019massistant}
Lan Jiang, Shuhan Hu, Mingyu Huang, Zhichun Wang, Jinjian Yang, Xiaoju Ye, and Wei Zheng.
\newblock Massistant: a personal knowledge assistant for mooc learners.
\newblock In \emph{Proceedings of the 2019 Conference on Empirical Methods in Natural Language Processing and the 9th International Joint Conference on Natural Language Processing (EMNLP-IJCNLP): System Demonstrations}, pages 133--138, 2019{\natexlab{a}}.

\bibitem[Jiang et~al.(2019{\natexlab{b}})Jiang, Pardos, and Wei]{jiang2019goal}
Weijie Jiang, Zachary~A Pardos, and Qiang Wei.
\newblock Goal-based course recommendation.
\newblock In \emph{Proceedings of the 9th international conference on learning analytics \& knowledge}, pages 36--45, 2019{\natexlab{b}}.

\bibitem[Jing and Tang(2017)]{jing2017guess}
Xia Jing and Jie Tang.
\newblock Guess you like: course recommendation in moocs.
\newblock In \emph{Proceedings of the international conference on web intelligence}, pages 783--789, 2017.

\bibitem[Kabudi et~al.(2021)Kabudi, Pappas, and Olsen]{kabudi2021ai}
Tumaini Kabudi, Ilias Pappas, and Dag~H{\aa}kon Olsen.
\newblock Ai-enabled adaptive learning systems: A systematic mapping of the literature.
\newblock \emph{Computers and Education: Artificial Intelligence}, 2:\penalty0 100017, 2021.

\bibitem[Koller and Ng(2013)]{koller2013online}
Daphne Koller and Andrew Ng.
\newblock The online revolution: Education for everyone.
\newblock In \emph{Seminar presentation at the Said Business School}. Oxford University England, 2013.

\bibitem[Kurdi et~al.(2020)Kurdi, Leo, Parsia, Sattler, and Al-Emari]{kurdi2020systematic}
Ghader Kurdi, Jared Leo, Bijan Parsia, Uli Sattler, and Salam Al-Emari.
\newblock A systematic review of automatic question generation for educational purposes.
\newblock \emph{International Journal of Artificial Intelligence in Education}, 30:\penalty0 121--204, 2020.

\bibitem[Nabizadeh et~al.(2020)Nabizadeh, Leal, Rafsanjani, and Shah]{nabizadeh2020learning}
Amir~Hossein Nabizadeh, Jos{\'e}~Paulo Leal, Hamed~N Rafsanjani, and Rajiv~Ratn Shah.
\newblock Learning path personalization and recommendation methods: A survey of the state-of-the-art.
\newblock \emph{Expert Systems with Applications}, 159:\penalty0 113596, 2020.

\bibitem[Nguyen et~al.(2023)Nguyen, Ngo, and Pham]{nguyen_automatic_nodate}
Nguyen Xuan~Vu Nguyen, Quang~Huy Ngo, and Quang Nhat~Minh Pham.
\newblock Automatic transcript generation from presentation slides.
\newblock In Chu-Ren Huang, Yasunari Harada, Jong-Bok Kim, Si~Chen, Yu-Yin Hsu, Emmanuele Chersoni, Pranav A, Winnie~Huiheng Zeng, Bo~Peng, Yuxi Li, and Junlin Li, editors, \emph{Proceedings of the 37th Pacific Asia Conference on Language, Information and Computation}, pages 670--678, Hong Kong, China, December 2023. Association for Computational Linguistics.
\newblock URL \url{https://aclanthology.org/2023.paclic-1.67}.

\bibitem[Nye et~al.(2014)Nye, Graesser, and Hu]{nye2014autotutor}
Benjamin~D Nye, Arthur~C Graesser, and Xiangen Hu.
\newblock Autotutor and family: A review of 17 years of natural language tutoring.
\newblock \emph{International Journal of Artificial Intelligence in Education}, 24:\penalty0 427--469, 2014.

\bibitem[Pal~Chowdhury et~al.(2024)Pal~Chowdhury, Zouhar, and Sachan]{chowdhury_autotutor_2024}
Sankalan Pal~Chowdhury, Vil\'{e}m Zouhar, and Mrinmaya Sachan.
\newblock Autotutor meets large language models: A language model tutor with rich pedagogy and guardrails.
\newblock In \emph{Proceedings of the Eleventh ACM Conference on Learning @ Scale}, L@S '24, page 5–15, New York, NY, USA, 2024. Association for Computing Machinery.
\newblock ISBN 9798400706332.
\newblock \doi{10.1145/3657604.3662041}.
\newblock URL \url{https://doi.org/10.1145/3657604.3662041}.

\bibitem[Papadakis(2023)]{papadakis2023moocs}
Stamatios Papadakis.
\newblock Moocs 2012-2022: An overview.
\newblock \emph{Advances in Mobile Learning Educational Research}, 3\penalty0 (1):\penalty0 682--693, 2023.

\bibitem[Pappano(2012)]{pappano2012year}
Laura Pappano.
\newblock The year of the mooc.
\newblock \emph{The New York Times}, 2\penalty0 (12):\penalty0 2012, 2012.

\bibitem[Park et~al.(2023{\natexlab{a}})Park, O'Brien, Cai, Morris, Liang, and Bernstein]{park2023generative}
Joon~Sung Park, Joseph O'Brien, Carrie~Jun Cai, Meredith~Ringel Morris, Percy Liang, and Michael~S Bernstein.
\newblock Generative agents: Interactive simulacra of human behavior.
\newblock In \emph{Proceedings of the 36th annual acm symposium on user interface software and technology}, pages 1--22, 2023{\natexlab{a}}.

\bibitem[Park et~al.(2023{\natexlab{b}})Park, O'Brien, Cai, Morris, Liang, and Bernstein]{Park2023GenerativeAgents}
Joon~Sung Park, Joseph~C. O'Brien, Carrie~J. Cai, Meredith~Ringel Morris, Percy Liang, and Michael~S. Bernstein.
\newblock Generative agents: Interactive simulacra of human behavior.
\newblock In \emph{In the 36th Annual ACM Symposium on User Interface Software and Technology (UIST '23)}, UIST '23, New York, NY, USA, 2023{\natexlab{b}}. Association for Computing Machinery.

\bibitem[Qian et~al.(2023)Qian, Cong, Yang, Chen, Su, Xu, Liu, and Sun]{qian2023communicative}
Chen Qian, Xin Cong, Cheng Yang, Weize Chen, Yusheng Su, Juyuan Xu, Zhiyuan Liu, and Maosong Sun.
\newblock Communicative agents for software development.
\newblock \emph{arXiv preprint arXiv:2307.07924}, 6, 2023.

\bibitem[Reich(2015)]{reich2015rebooting}
Justin Reich.
\newblock Rebooting mooc research.
\newblock \emph{Science}, 347\penalty0 (6217):\penalty0 34--35, 2015.

\bibitem[Reich and Ruip{\'e}rez-Valiente(2019)]{reich2019mooc}
Justin Reich and Jos{\'e}~A Ruip{\'e}rez-Valiente.
\newblock The mooc pivot.
\newblock \emph{Science}, 363\penalty0 (6423):\penalty0 130--131, 2019.

\bibitem[Rizvi et~al.(2022)Rizvi, Rienties, Rogaten, and Kizilcec]{rizvi2022beyond}
Saman Rizvi, Bart Rienties, Jekaterina Rogaten, and Ren{\'e}~F Kizilcec.
\newblock Beyond one-size-fits-all in moocs: Variation in learning design and persistence of learners in different cultural and socioeconomic contexts.
\newblock \emph{Computers in Human Behavior}, 126:\penalty0 106973, 2022.

\bibitem[Schick et~al.(2024)Schick, Dwivedi-Yu, Dess{\`\i}, Raileanu, Lomeli, Hambro, Zettlemoyer, Cancedda, and Scialom]{schick2024toolformer}
Timo Schick, Jane Dwivedi-Yu, Roberto Dess{\`\i}, Roberta Raileanu, Maria Lomeli, Eric Hambro, Luke Zettlemoyer, Nicola Cancedda, and Thomas Scialom.
\newblock Toolformer: Language models can teach themselves to use tools.
\newblock \emph{Advances in Neural Information Processing Systems}, 36, 2024.

\bibitem[Schwanke(1981)]{schwanke1981classroom}
Dean Schwanke.
\newblock Classroom interaction research: A survey of recent literature.
\newblock \emph{Journal of Classroom Interaction}, pages 8--10, 1981.

\bibitem[Thornberg et~al.(2022)Thornberg, Forsberg, Hammar~Chiriac, and Bjereld]{thornberg2022teacher}
Robert Thornberg, Camilla Forsberg, Eva Hammar~Chiriac, and Ylva Bjereld.
\newblock Teacher--student relationship quality and student engagement: A sequential explanatory mixed-methods study.
\newblock \emph{Research papers in education}, 37\penalty0 (6):\penalty0 840--859, 2022.

\bibitem[Touvron et~al.(2023)Touvron, Lavril, Izacard, Martinet, Lachaux, Lacroix, Rozi{\`e}re, Goyal, Hambro, Azhar, et~al.]{touvron2023llama}
Hugo Touvron, Thibaut Lavril, Gautier Izacard, Xavier Martinet, Marie-Anne Lachaux, Timoth{\'e}e Lacroix, Baptiste Rozi{\`e}re, Naman Goyal, Eric Hambro, Faisal Azhar, et~al.
\newblock Llama: Open and efficient foundation language models.
\newblock \emph{arXiv preprint arXiv:2302.13971}, 2023.

\bibitem[Tu et~al.(2023)Tu, Zhang, Yu, Li, Zhang, Yao, Hou, and Li]{tu2023littlemu}
Shangqing Tu, Zheyuan Zhang, Jifan Yu, Chunyang Li, Siyu Zhang, Zijun Yao, Lei Hou, and Juanzi Li.
\newblock Littlemu: Deploying an online virtual teaching assistant via heterogeneous sources integration and chain of teach prompts.
\newblock In \emph{Proceedings of the 32nd ACM International Conference on Information and Knowledge Management}, pages 4843--4849, 2023.

\bibitem[Wu et~al.(2023)Wu, Bansal, Zhang, Wu, Zhang, Zhu, Li, Jiang, Zhang, and Wang]{wu2023autogen}
Qingyun Wu, Gagan Bansal, Jieyu Zhang, Yiran Wu, Shaokun Zhang, Erkang Zhu, Beibin Li, Li~Jiang, Xiaoyun Zhang, and Chi Wang.
\newblock Autogen: Enabling next-gen llm applications via multi-agent conversation framework.
\newblock \emph{arXiv preprint arXiv:2308.08155}, 2023.

\bibitem[Xie(2023)]{xie2023influence}
Xiaofang Xie.
\newblock Influence of ai-driven inquiry teaching on learning outcomes.
\newblock \emph{International Journal of Emerging Technologies in Learning}, 18\penalty0 (23), 2023.

\bibitem[Yilmaz et~al.(2022)Yilmaz, Yurdugul, Yilmaz, Sahin, Sulak, Aydin, Tepgec, Muftuoglu, and Omer]{yilmaz2022smart}
Ramazan Yilmaz, Halil Yurdugul, Fatma Gizem~Karaoglan Yilmaz, Muhittin Sahin, Sema Sulak, Furkan Aydin, Mustafa Tepgec, Cennet~Terzi Muftuoglu, and ORAL Omer.
\newblock Smart mooc integrated with intelligent tutoring: A system architecture and framework model proposal.
\newblock \emph{Computers and Education: Artificial Intelligence}, 3:\penalty0 100092, 2022.

\bibitem[Yu et~al.(2020)Yu, Luo, Xiao, Zhong, Wang, Feng, Luo, Wang, Hou, Li, et~al.]{yu2020mooccube}
Jifan Yu, Gan Luo, Tong Xiao, Qingyang Zhong, Yuquan Wang, Wenzheng Feng, Junyi Luo, Chenyu Wang, Lei Hou, Juanzi Li, et~al.
\newblock Mooccube: A large-scale data repository for nlp applications in moocs.
\newblock In \emph{Proceedings of the 58th annual meeting of the association for computational linguistics}, pages 3135--3142, 2020.

\bibitem[Yu et~al.(2021)Yu, Wang, Zhong, Luo, Mao, Sun, Feng, Xu, Cao, Zeng, et~al.]{yu2021mooccubex}
Jifan Yu, Yuquan Wang, Qingyang Zhong, Gan Luo, Yiming Mao, Kai Sun, Wenzheng Feng, Wei Xu, Shulin Cao, Kaisheng Zeng, et~al.
\newblock Mooccubex: a large knowledge-centered repository for adaptive learning in moocs.
\newblock In \emph{Proceedings of the 30th ACM International Conference on Information \& Knowledge Management}, pages 4643--4652, 2021.

\bibitem[Yue et~al.(2024)Yue, Mifdal, Zhang, Suh, and Yao]{yue2024mathvc}
Murong Yue, Wijdane Mifdal, Yixuan Zhang, Jennifer Suh, and Ziyu Yao.
\newblock Mathvc: An llm-simulated multi-character virtual classroom for mathematics education.
\newblock \emph{arXiv preprint arXiv:2404.06711}, 2024.

\bibitem[Zhang et~al.(2018)Zhang, Huang, Lv, Liu, and Zhou]{zhang2018mcrs}
Hao Zhang, Tao Huang, Zhihan Lv, SanYa Liu, and Zhili Zhou.
\newblock Mcrs: A course recommendation system for moocs.
\newblock \emph{Multimedia Tools and Applications}, 77:\penalty0 7051--7069, 2018.

\bibitem[Zhang et~al.(2024)Zhang, Zhang-Li, Yu, Gong, Zhou, Liu, Hou, and Li]{zhang2024simulating}
Zheyuan Zhang, Daniel Zhang-Li, Jifan Yu, Linlu Gong, Jinchang Zhou, Zhiyuan Liu, Lei Hou, and Juanzi Li.
\newblock Simulating classroom education with llm-empowered agents.
\newblock \emph{arXiv preprint arXiv:2406.19226}, 2024.

\bibitem[Zheng et~al.(2024)Zheng, Li, Huang, Liang, Guo, Hou, Gao, Tian, Liu, and Luo]{olney_automatic_2024}
Ying Zheng, Xueyi Li, Yaying Huang, Qianru Liang, Teng Guo, Mingliang Hou, Boyu Gao, Mi~Tian, Zitao Liu, and Weiqi Luo.
\newblock Automatic {Lesson} {Plan} {Generation} via {Large} {Language} {Models} with {Self}-critique {Prompting}.
\newblock In Andrew~M. Olney, Irene-Angelica Chounta, Zitao Liu, Olga~C. Santos, and Ig~Ibert Bittencourt, editors, \emph{Artificial {Intelligence} in {Education}. {Posters} and {Late} {Breaking} {Results}, {Workshops} and {Tutorials}, {Industry} and {Innovation} {Tracks}, {Practitioners}, {Doctoral} {Consortium} and {Blue} {Sky}}, volume 2150, pages 163--178. Springer Nature Switzerland, Cham, 2024.
\newblock ISBN 978-3-031-64314-9 978-3-031-64315-6.
\newblock \doi{10.1007/978-3-031-64315-6_13}.
\newblock URL \url{https://link.springer.com/10.1007/978-3-031-64315-6_13}.
\newblock Series Title: Communications in Computer and Information Science.

\bibitem[Zhong et~al.(2022)Zhong, Yu, Zhang, Mao, Wang, Lin, Hou, Li, and Tang]{zhong2022towards}
Qingyang Zhong, Jifan Yu, Zheyuan Zhang, Yiming Mao, Yuquan Wang, Yankai Lin, Lei Hou, Juanzi Li, and Jie Tang.
\newblock Towards a general pre-training framework for adaptive learning in moocs.
\newblock \emph{arXiv preprint arXiv:2208.04708}, 2022.

\bibitem[Zhu et~al.(2020)Zhu, Sari, and Lee]{zhu2020comprehensive}
Meina Zhu, Annisa~R Sari, and Mimi~Miyoung Lee.
\newblock A comprehensive systematic review of mooc research: Research techniques, topics, and trends from 2009 to 2019.
\newblock \emph{Educational Technology Research and Development}, 68:\penalty0 1685--1710, 2020.

\end{thebibliography}

%%%%%%%%%%%%%%%%%%%%%%%%%%%%%%%%%%%%%%%%%%%%%%%%%%%%%%%%%%%%

% \appendix

% \section{Appendix}

\end{document}